\pdfoutput=1

\documentclass[rmp,aps,reprint]{revtex4-2}
\usepackage{graphicx}% Include figure files
\usepackage{subfigure}
\usepackage{dcolumn}% Align table columns on decimal point
\usepackage{bm}% bold math
\usepackage{textcomp}
\usepackage{gensymb}
\usepackage{amsmath}
\usepackage{amssymb}
\usepackage{dsfont}
\usepackage{hyperref}
\hypersetup{
    colorlinks=true,
    linkcolor=blue,
    filecolor=magenta,      
    urlcolor=cyan,
    }
\usepackage{multirow}
\usepackage{rotating}
\usepackage{color}
\usepackage{fancyhdr}
\usepackage{lipsum,tikz}
\usepackage{tabularx, booktabs}
\usepackage{etoolbox}
\usepackage[strings]{underscore}
\usepackage{mathpazo}
\usepackage{mathtools}
\usepackage{physics}

\begin{abstract} % abstract
We take a qualitative comparative look at quantum and classical quartic anharmonic oscillators. It has been shown that the behavior of the quantum anharmonic oscillator mimics that of the classical anharmonic oscillators with the structurally same Hamiltonian in the coherent state basis of the harmonic oscillators. The associated equation of motion allows us to use Lindstet-Poincare perturbation method to compute the classical frequency of the oscillation order by order, by taking care of its dependence on amplitude and the quantum corrections. We also derive a bound for periodicity of such oscillations in both the classical and quantum cases.
\end{abstract}

\begin{document}
		
\title{Quantum-Classical Correspondence in a Quartic Oscillator - Corrections to Frequency and Bounds for Periodic Motion}

\author{Mandas~Biswas$^1$}
\author{Deb Shankar Ray$^2$}
\newcommand{\IACS}{\affiliation{$^1$School of Physical Sciences, Indian Association for the Cultivation of Science, Jadavpur, Kolkata 700032, India.
\\$^2$School of Chemical Sciences, Indian Association for the Cultivation of Science, Jadavpur, Kolkata 700032, India.}}
\IACS

\maketitle

\tableofcontents

\section{Introduction}
Anharmonic oscillators are ubiquitous in natural sciences. They form the basic paradigm for classical oscillators and quantum vibrations in optical systems, lattice and molecular vibration, acoustics and in general. In contrast to linear oscillators, their exact solutions are unfeasible. A simple way out of the difficulty is to take resort to perturbation theory, which has found wide applicability, e.g., in finding out the linear and nonlinear response of a system to an external stimulus. The development in this and allied areas have contributed a  large body of literature over many decades, \cite{10.1119/1.3579129}\cite{Akande22}
\\One of the major difficulties of the traditional perturbation theory, when applied to a driven or nonlinear system is that the results are valid for short time and for weak stimulus, since most often the perturbation series diverges because of the appearance of the secular terms, \cite{Schro} For example, Dirac perturbation series in time-dependent theory is inherently divergent although its usefulness for calculations of the probability per unit time  in the linear regime has long been well-accepted because of its phenomenal success in explaining light-matter interaction. In what follows, we revisit the issue of divergence in the context of classical and quantum version of a nonlinear oscillator from the perspective of oscillatory motion in the steady state and its bound. It is well-known that a divergence-free perturbation theory (like Lindstet-Poincare\cite{article}) is more appropriate that captures the correct frequency of oscillation by taking care of its dependence on the amplitude of motion. The object of the present work is to explore this amplitude dependence of frequency in classical and quantum regime by constructing an equation of motion of a quantum quartic oscillator in the coherent state basis\cite{doi:10.1142/0096} of harmonic oscillator. In principle, we could have used the coherent states of this Hamiltonian itself, but that is a non-trivial issue in itself, as showcased in the article, \cite{sharatchandra1997coherentstatesanharmonicoscillator}. We calculate the classical frequency and its quantum corrections order-by-order with appropriate amplitude dependence. An interesting consequence of the analysis is the realization of the condition for isochronicity\cite{article} of the nonlinear oscillator, i.e., the condition for which the oscillator becomes independent of the amplitude for specific values of the nonlinear coefficient. We also analyze the motion along the seperatrix to demonstrate how quantum corrections alter the classical bounds on the amplitude.
\section{Equation of motion of a Quartic Oscillator in a Coherent Basis}
We consider the quantum Hamiltonian for a quartic potential term in one dimension,
\begin{equation}
    \hat{H} = \frac{\hat{p^2}}{2m} + \frac{1}{2}m\omega^2\hat{x^2} + \lambda\hat{x^4}
\end{equation}
From now onwards, the hat is dropped for brevity. Here, $\omega$ is the linear frequency, and $\lambda$ is the nonlinear coefficient. In terms of the annihilation and creation operators (given by $a = \sqrt{\frac{m\omega}{2\hbar}}(x + \frac{i}{m\omega}p)$, and its hermitian conjugate $a^\dagger$, with x and p as being the position and momentum operator respectively), the Hamiltonian may be expressed as, 
\begin{equation}
    H = \hbar\omega(a^\dagger a + 1/2\hat{I}) + \lambda[\sqrt{\frac{\hbar}{2m\omega}}(a + a^\dagger)]^4 = H_0 + H_I
\end{equation}
where, $H_0 = \hbar\omega(a^\dagger a + 1/2\hat{I})$, and, $H_I = \frac{\lambda{\hbar}^2}{(2m\omega)^2}(a^4 + 4a^3a^\dagger + 6a^2{a^\dagger}^2 + 4a{a^\dagger}^3 + {a^\dagger}^4)$. $a$ and $a^\dagger$ follow the boson commutation rule, $[a, a^\dagger] = 1$. 
\\We know that, coherent states ($\ket{\alpha}$) of a quantum harmonic oscillator are the eigenstates of the lowering operator, given by $a$, and their eigenvalues are generally complex. They follow,
\begin{equation}
    a\ket{\alpha} = \alpha\ket{\alpha}; \ket{\alpha} = e^{-|\alpha|^2/2}\Sigma_{n=0}^{\infty} \frac{\alpha^n}{\sqrt{n!}}\ket{n}
\end{equation}
where, $\ket{n}$ is the number state, that is, the energy eigenbasis written in Fock convention. Any intended coherent states can be created out of the ground energy eigenbasis $\ket{0}$ with the aid of a displacement operator, 
\begin{equation}
    \ket{\alpha} = D(\alpha)\ket{0} = e^{-|\alpha|^2/2}e^{\alpha a^\dagger}e^{-\alpha^* a}\ket{0}
\end{equation}
The expecation value of the postion and momentum operators in the coherent state is given by,
\begin{equation}
    <x>_\alpha = \sqrt{\frac{\hbar}{2m\omega}}(\alpha + \alpha^*) \text{Or, } k_1<x>_\alpha = (\alpha + \alpha^*)
\end{equation}
\begin{equation}
    <p>_\alpha = -i\sqrt{\frac{m\omega\hbar}{2}}(\alpha - \alpha^*) \text{Or, } k_2<p>_\alpha = (\alpha - \alpha^*)
\end{equation}
Here, the values of the constants are, $k_1 = \sqrt{\frac{2m\omega}{\hbar}}$, and $k_2 = \sqrt{i\frac{2}{m\omega\hbar}}$. 
Using these, we have the Heisenberg equations of motion,
\begin{eqnarray}
    i\hbar\frac{d}{dt}a &=& [a, H]\\
    \text{Or, } i\hbar\frac{d}{dt}\bra{\alpha}a\ket{\alpha} &=& \bra{\alpha}[a, H]\ket{\alpha}\nonumber\\
    &=& \bra{\alpha}\hbar\omega[a, (a^\dagger a)] + \lambda[a, x^4]\ket{\alpha}
\end{eqnarray}
Working out explicitly the expectation values of the above commutations in the coherent basis, we obtain\footnote{Though it's not required to carry out the commutation if one is careful enough, these commutation relations may come in handy sometimes, $[a, (a^\dagger)^n] = n(a^\dagger)^{n-1}$, and, $[a^\dagger, a^n] = -na^{n-1}$. Their proofs are easy to see by induction.}, 
\begin{eqnarray}
    \bra{\alpha}[a, (a^\dagger a)]\ket{\alpha} &=& \bra{\alpha}aa^\dagger a\ket{\alpha} - \bra{\alpha}a^\dagger a^2\ket{\alpha}\nonumber\\
    &=& \alpha(1 + \alpha^*\alpha) - \alpha^2\alpha^*\nonumber\\
    &=& \alpha
\end{eqnarray}
And, we also find, 
\begin{eqnarray}
    \bra{\alpha}\lambda[a, x^4]\ket{\alpha} &=& \frac{\lambda\hbar^2}{(2m\omega)^2}[4\alpha^3 + 4(\alpha^*)^3 + 24\alpha + 24\alpha^* \nonumber\\
    &+& 12\alpha^*\alpha^2 + 12(\alpha^*)^2\alpha]
\end{eqnarray}
Using the relations (8), (9) and (10), we obtain from (7), 
\begin{equation}
    i\hbar\frac{d}{dt}<a>_\alpha = \hbar\omega\alpha + \frac{\lambda\hbar^2}{(2m\omega)^2}[4k_1^3<x>_\alpha^3 + 24k_1<x>_\alpha]
\end{equation}
and,
\begin{equation}
    i\hbar\frac{d}{dt}<a^\dagger>_\alpha = -\hbar\omega\alpha^* - \frac{\lambda\hbar^2}{(2m\omega)^2}[4k_1^3<x>_\alpha^3 + 24k_1<x>_\alpha]
\end{equation}
Eqns. (11) and (12) give us,
\begin{eqnarray}
    i\hbar\frac{k_1}{k_2}\dot{x} &=& \hbar\omega p\\
    i\hbar\frac{k_2}{k_1}\dot{p} &=& \hbar\omega x + \frac{\lambda\hbar^2}{2m^2\omega^2}[4k_1^2x^3 + 24x]
\end{eqnarray}
From eqns. (13) and (14), it is straightforward to obtain,
\begin{equation}
    \ddot{x} + \frac{\omega^2}{\hbar}x + \frac{\lambda}{2m^2\omega}(4k_1^2x^3 + 24x) = 0
\end{equation}
This is the equation of motion of the quantum anharmonic oscillator in the coherent state basis of a quantum harmonic oscillator.
\\On the other hand, the Hamiltonian in the classical picture is,
\begin{equation}
    H = \frac{p^2}{2m} + \frac{1}{2}m\omega^2 x^2 + \lambda x^4
\end{equation}
The Hamilton's equations of motions are given by,
\begin{eqnarray}
    \dot{x} &=& \frac{p}{m}\\
    \dot{p} &=& -m\omega^2 x - 4\lambda x^3
\end{eqnarray}
Or,
\begin{equation}
    \ddot{x} + \omega^2 x + \frac{4\lambda}{m}x^3 = 0
\end{equation}
This is the equation of motion of classical anharmonic oscillator.
\section{Solutions for Oscillatory and Seperatrix Motion}
It is very easy to check that both equations (15) and (19) can be written in a common form as follows,
\begin{equation}
    \ddot{x} + \epsilon_1 x + \epsilon_2 x^3 = 0
\end{equation}
where, in the QM case, $\epsilon_1 = 4k_1^2\times \frac{\lambda\hbar^2}{2m^2\omega^2} = \frac{4\lambda}{m\hbar}$, and $\epsilon_2 = \frac{\omega^2}{\hbar} + \frac{12\lambda}{m^2\omega}$; whereas, in the CM case, we have, $\epsilon_1 = \omega^2$ and $\epsilon_2 = \frac{4\lambda}{m}$. In what follows, we explore the equations in the context of a quantum-classical correspondence, in this section. Two pertinent points are to be noted. First, because of nonlinearity in eq. (10), the application of traditional perturbation method leads to divergence. To bypass this difficulty, we employ Lindstet-Poincare method which correctly takes care of the amplitude dependence of frequency of a nonlinear oscillator. An interesting offshoot of this calculation is the condition of the removal of secular divergence, which, in turn, results in a condition for isochronicity of the oscillator for specific values of nonlinear coeffcient. Second, depending on the sign of $\epsilon_1$ and $\epsilon_2$, the oscillator can be of a Duffing type ($\epsilon_1 > 0$ and $\epsilon_2 >0$), double well-type ($\epsilon_1<0$ and $\epsilon_2>0$) or inverted double-well type ($\epsilon_1 > 0$ and $\epsilon_2 <0$). For the first case, we have only one steady state (0, 0), whereas, for the other two cases we have three steady states. For the double-well the unstable fixed point lies at the barrier top which separates the the small oscillations in the two wells centering at the two stable fixed points located at the bottom of the two wells. A separatrix line separates out the large oscillations above the barrier from small oscillations in the well. Our next aim is to calculate the time period of the round trip motion along the separatrix and look for the segments of quantum corrections.   
\subsection{Divergence-Free Perturbative Solution; Condtion for Isochronicity}
It is convenient to change the time-scale of the equation (20) by $\tau = \sqrt{\epsilon_1}t$.
\\So, we have,
\begin{equation}
    \frac{d^2x}{d\tau^2} + x + \frac{\epsilon_2}{\epsilon_1}x^3 = 0
\end{equation}
Let, $\frac{\epsilon_2}{\epsilon_1} = b$, for brevity. b can be used as a smallness parameter since, $b<<1$ in both cases. We take, $b>0$.  
\\This is exactly in the form of a \textit{Duffing oscillator} equation. We introduce the dimensionless time, $T = \Omega \tau$, where, $\Omega$ is to be determined. Substituting this in eq. (21), we get,
\begin{equation}
    \Omega^2\frac{d^2x}{dT^2} + x + bx^3 = 0
\end{equation}
It is interesting to note that if b=0, then the equation actually reduces to the equation of a simple harmonic oscillator equation. Using a power series expansion of both x and $\Omega$ in terms of b to get the solution of the equation (21),
\begin{eqnarray}
    \Omega &=& 1 + \Omega_1b + \Omega_2b^2 + ...\\
    x(T) &=& x_0(T) + bx_1(T) + b^2x_2(T) + ...
\end{eqnarray}
Substituting (23) and (24) in eq. (22) we collect terms in orders of b to obtain,
\begin{eqnarray}
    \frac{d^2x_0}{dT^2} + x_0 &=& 0\\
    \frac{d^2x_1}{dT^2} + x_1 &=& -2\Omega_1\frac{d^2x_0}{dT^2} - x_0^3\\
    \frac{d^2x_2}{dT^2} + x_2 &=& -2\Omega_1\frac{d^2x_1}{dT^2} - (2\Omega_2 + \Omega_1^2)\frac{d^2x_0}{dT^2}\nonumber\\
    &-& 3x_0^3x_1
\end{eqnarray}
Equation (25) has the solution, $x_0(T) = A\cos T$, where, A is the amplitude of the motion and the phase is chosen arbitrarily because of the autonomous nature of equation (21). Substituting this solution in eq. (26), we have,
\begin{equation}
    \frac{d^2x_1}{dT^2} + x_1 = 2A\Omega_1\cos T - A^3\cos^3T
\end{equation}
Simplifying the $\cos^3T$ term, we get,
\begin{equation}
    \frac{d^2x_1}{dT^2} + x_1 = (2A\Omega_1 - \frac{3A^3}{4})\cos T - \frac{A^3}{4}\cos3T
\end{equation}
For a periodic solution, we require the coefficient of $\cos T$ on the right hand side of eq.(29) to vanish. This key step is the condition of removal of resonance or secular terms. Using this, we obtain the first order correction to the frequency,
\begin{equation}
    \Omega_1 = \frac{3}{8}A^2
\end{equation}
So, substituting eq. (30) in eq. (23), we have, till first order of b, the frequency of the motion of the system, 
\begin{equation}
    \Omega = 1 + \frac{3}{8}A^2b
\end{equation}
As we can clearly see, the frequency is dependent on the amplitude of the motion. Now the nature of quantum and classical motion depends on b. To express this, we note that, 
\begin{equation}
    b)_{CM} = \frac{4\lambda}{m\omega^2}
\end{equation}
For QM, this turns out to be,
\begin{eqnarray}
    b)_{QM} &=& \frac{4\lambda}{m\hbar}(\frac{\omega^2}{\hbar} + \frac{12\lambda}{m^2\omega})^{-1}\nonumber\\
    &=& \frac{4\lambda}{m\omega^2}(1 + \frac{12\lambda\hbar}{m^2\omega^3})^{-1}\nonumber\\
    &=& \frac{4\lambda}{m\omega^2}(1 - \frac{12\lambda\hbar}{m^2\omega^3})\nonumber\\
    \text{Or, } b)_{QM} &=& \frac{4\lambda}{m\omega^2} - \frac{48\lambda^2\hbar}{m^3\omega^5}
\end{eqnarray}
where, we have kept terms upto $O(\hbar)$. From the eq. (30), it is revealed that, the frequency of the system,
\begin{eqnarray}
    \Omega)_{CM} &=& 1 + \frac{3A^2\lambda}{2m\omega^2}\\
    \Omega)_{QM} &=& 1 + \frac{3A^2\lambda}{2m\omega^2} - \frac{18\lambda^2A^2\hbar}{m^3\omega^5}
\end{eqnarray}
It is now apparent that the second term on the RHS of the equation (35) is precisely a quantum correction to the frequency of the system, which was absent when we treated the system in a classical fashion. Now, we note that for any arbitrary value of $\lambda$ the frequency of the classical motion is always dependent on the amplitude. The quantum correction in (35), however, makes the quantum frequency lower than the classical one. Because of the negativity, the quantum correction may vanish for specific choice of $\lambda$; $\lambda = \frac{m^2\omega^3}{12\hbar}$. Under the condition, $\Omega = 1$, i.e., the quantum oscillator does not depend on the amplitude of motion. Hence, the quantum motion is isochronous in contrast to the classical motion.   
\\We may continue the process to obtain higher order approximations to check whether this quantum isochronicity is only a first order semiclassical effect, as follows. Substituting condition (30) into eq.(29), we may solve for $x_1(T)$,
\begin{equation}
    x_1(T) = \frac{A^3}{32}(\cos3T - \cos T)
\end{equation}
Here we have chosen the complementary solution in order that the amplitude of vibration be given by A, that is, in order that x(0) = A. Substituting (36) into the $x_2$ equation (33), we may again remove the secular terms and thereby obtain an expression for $\Omega_2$. The process may be continued indefinitely.
\\So let's check for $O(b^2)$. If we take $x_0$ and $x_1$ and substitute in (27), we get,
\begin{equation}
    \frac{d^2x_2}{dT^2} + x_2 = -2\Omega_1\frac{d^2x_1}{dT^2} - (2\Omega_2 + \Omega_1^2)\frac{d^2x_0}{dT^2}
\end{equation}
We put in the value of $\Omega_1$. Then,
\begin{eqnarray}
    \frac{d^2x_2}{dT^2} + x_2 &=& \frac{A}{128}(21A^4 + 256\Omega_2)\cos T + \frac{24A^5}{128}\cos 3T\nonumber\\
    &-& \frac{3A^5}{128}\cos 5T
\end{eqnarray}
To remove secular terms, we set coefficient of $\cos T$ to be zero, that is,
\begin{equation}
    \Omega_2 = -\frac{21A^4}{256}
\end{equation}
With,
\begin{equation}
    b^2)_{CM} = \frac{16\lambda^2}{m^2\omega^4}
\end{equation}
and,
\begin{eqnarray}
    b^2)_{QM} &=& \frac{16\lambda^2}{m^2\hbar^2}(\frac{\omega^2}{\hbar} + \frac{12\lambda}{m^2 \omega})^{-2} \nonumber\\
    &=& \frac{16\lambda^2}{m^2\omega^4}(1 - \frac{24\lambda\hbar}{m^2\omega^3})\nonumber\\
    &=& \frac{16\lambda^2}{m^2\omega^4} - \frac{384\lambda^3\hbar}{m^4\omega^7}
\end{eqnarray}
We have,
\begin{equation}
    \Omega = 1 + \frac{3}{8}A^2b - \frac{21A^4}{256}b^2
\end{equation}
So, for the CM case, we are led to,
\begin{equation}
    \Omega)_{CM} = 1 + \frac{3A^2\lambda}{2m\omega^2} - \frac{21A^4\lambda^2}{16m^2\omega^4}
\end{equation}
Therefore, the CM case frequency becomes isochronous (considering that $\lambda \neq 0$), if $\lambda = \frac{8m\omega^2}{7A^2}$.
\\For the QM case, we have, on the other hand,
\begin{eqnarray}
    \Omega)_{QM} &=& 1 + \frac{3A^2\lambda}{2m\omega^2} - \frac{18\lambda^2A^2\hbar}{m^3\omega^5}\nonumber\\ 
    &-& \frac{21A^4}{256}(\frac{16\lambda^2}{m^2\omega^4} - \frac{384\lambda^3\hbar}{m^4\omega^7})\nonumber\\
    &=& 1 + \frac{3A^2\lambda}{2m\omega^2} - \frac{18\lambda^2A^2\hbar}{m^3\omega^5} - \frac{21A^4\lambda^2}{16m^2\omega^4}\nonumber\\
    &+& \frac{63A^4\lambda^3\hbar}{2m^4\omega^7}
\end{eqnarray}
So, if the oscillator is isochronous, then, for some value of $\lambda$, we must have,
\begin{equation}
    \frac{3A^2\lambda}{2m\omega^2} - \frac{18\lambda^2A^2\hbar}{m^3\omega^5} - \frac{21A^4\lambda^2}{16m^2\omega^4} + \frac{63A^4\lambda^3\hbar}{2m^4\omega^7} = 0
\end{equation}
For $\lambda\neq 0$, the real solutions exist for,
\begin{equation}
    (\frac{18A^2\hbar}{m^3\omega^5} + \frac{21A^4}{16m^2\omega^4}) - \frac{189A^6\hbar}{m^5\omega^9} \geq 0
\end{equation}
Thereby, we can conclude that this isochronous quantum oscillation of an anharmonic oscillator is presumably a result of the interference of non-linearity ($\lambda$) and quantum noise ($\hbar$) and a key aspect of this study. 
\subsection{The motion along Separatrix; Quantum Corrections for the Bounds of Oscillations}
It is also interesting to solve the system for the motion along the seperatrix for both inverted and vertical double-well potential, that is, $\lambda \longrightarrow -\lambda$ and $\omega \longrightarrow -\omega$. 
\subsubsection{Inverted Double-Well Potential}
For, $\lambda$ to $-\lambda$, the Hamiltonian becomes,
\begin{equation}
    H = \frac{p^2}{2m} + \frac{1}{2}m\omega^2x^2 - \lambda x^4
\end{equation}
It is easy to see that, the corresponding equation of motion turns out to be,
\begin{equation}
    \frac{d^2x}{d\tau^2} + x - bx^3 = 0
\end{equation}
Since, $\frac{dx}{d\tau} = p$, and, $\frac{dp}{d\tau} = -x + bx^3$, we have,
\begin{eqnarray}
    \frac{dp}{dx} &=& \frac{-x+bx^3}{p}\nonumber\\
    \text{Or, } \frac{p^2}{2} + \frac{x^2}{2} - \frac{bx^4}{4} &=& C
\end{eqnarray}
C is an integration constant, the total energy of the system (the system has time-translation symmetry).
\\This is the phase space equation of the system. It has got two nontrivial equilibrium points (p = $\dot{x}$ = 0) - $x = \pm\frac{1}{\sqrt{b}}, 0, \text{and, for all the x-values } y=0.$ The integral curves which go through these points separate motions which are periodic from motions which grow unbounded, and are identified as separatrices. These two fixed points except for the origin are saddle points to the curve, and hence are unstable equilibrium points. 
\\For the motion with, C=0, we rearrange the terms a bit to get,
\begin{eqnarray}
    p &=& x\sqrt{\frac{bx^2}{2} - 1}\nonumber\\
    \text{Or, } \int^{T/2}_{0} d\tau &=& \int^{A}_{0} \frac{dx}{x\sqrt{\frac{bx^2}{2} - 1}}
\end{eqnarray}
Substituting $\frac{bx^2}{2} = \sec^2\theta$, so, one obtains, $\theta = \cos^{-1}(\sqrt{\frac{2}{b}}\frac{1}{x})$ and $dx = \sqrt{\frac{2}{b}}\sec\theta\tan\theta d\theta$. Finally, we have,
\begin{eqnarray}
    \frac{T}{2} &=& \int^{A}_{0}d\theta\nonumber\\
    \frac{T}{2} &=& \sec^{-1}(\sqrt{\frac{b}{2}}A) - \sec^{-1}0
\end{eqnarray}
So, the integral becomes undefined. It is  expected, because C=0 corresponds to the equation of motion being the separatrix itself, and thus, in principle being a periodic motion but the time period tending to infinity. It is supported by our finding, and also - we note that the time period is directly proportional to the amplitude of the motion too.
\\Now, let C is positive, let it be E. We will have, proceeding in a similar fashion,
\begin{eqnarray}
    \frac{dx}{d\tau} &=& \sqrt{\frac{bx^4}{2} - x^2 + 2E}\nonumber\\
    \text{Or, } \int^{T/2}_{0} d\tau &=& \int^{A}_{0} \frac{dx}{\sqrt{\frac{bx^4}{2} - x^2 + 2E}}\nonumber\\
    \text{Or, } \frac{T}{2} &=& \int^{A}_{0} \frac{dx}{x\sqrt{\frac{bx^2}{2} - 1 + \frac{2E}{x^2}}}
\end{eqnarray}
To proceed furher, we consider the following situation, for which we choose E such that, let, {$2\sqrt{bE} = 1$, a \textbf{necessary} condition that is used. Under this condition, $\frac{bx^2}{2} - 1 + \frac{2E}{x^2} = [\sqrt{\frac{b}{2}}x - \frac{\sqrt{2E}}{x}]^2$. With this, (52) becomes,
\begin{equation}
    \frac{T}{2} = \frac{1}{\sqrt{2E}}\int^{A}_{0} \frac{dx}{\sqrt{\frac{b}{4E}}x^2 - 1}
\end{equation}
The value of this integral (given that $\tanh^{-1}0 = 0$) can be found out to be,
\begin{equation}
    T = -\sqrt{\frac{2}{E}}\frac{\tanh^{-1}((\frac{b}{4E})^{1/4}A)}{(\frac{b}{4E})^{1/4}}
\end{equation}
Since the argument of $\tanh^{-1}x$ is always greater than zero and has to be lesser than one, and, given the condition,
\begin{equation}
    A < \frac{1}{\sqrt{b}}
\end{equation}
It must hold if T has a positive solution.
\\Again, it is straightforward to obtain $\tanh^{-1}\gamma = \frac{1}{2}\ln(\frac{1+\gamma}{1-\gamma})$. If the time period of the motion has to be positive, the argument of the natural logarithm has to be lesser than e, that is,
\begin{eqnarray}
    \frac{1+x}{1-x} &<& e\nonumber\\
    \text{Or, } x &<& \frac{e-1}{e+1} \approx 0.46211715726 = k (\text{say})
\end{eqnarray}
Therefore, we get a bound for the time period of the motion to be solvable in terms of the amplitude of the motion for the condition,
\begin{equation}
    A < \frac{k}{\sqrt{b}}
\end{equation}
Putting the values of b for the classical and quantum cases, we have,
\begin{equation}
    A_{CM} < \frac{k\omega}{2}\sqrt{\frac{m}{\lambda}}
\end{equation}
and,
\begin{equation}
    A_{QM} < \frac{k}{\sqrt{\frac{4\lambda}{m\omega^2} - \frac{48\lambda^2\hbar}{m^3\omega^5}}}
\end{equation}
\subsubsection{Double-Well Potential}
For this case, the Hamiltonian is,
\begin{equation}
    H = \frac{p^2}{2m} - \frac{1}{2}m\omega^2x^2 + \lambda x^4
\end{equation}
That implies that the corresponding equation of motion is,
\begin{equation}
    \ddot{x} - \omega^2 x + \frac{4\lambda}{m}x^3 = 0
\end{equation}
Changing to $\tau$, then, we have,
\begin{equation}
    \frac{d^2x}{d\tau^2} - x + bx^3 = 0
\end{equation}
Again, we have the constant of motion C as given by,
\begin{eqnarray}
 \frac{p^2}{2} - \frac{x^2}{2} + \frac{bx^4}{4} &=& C
\end{eqnarray}
For C = 0, we have,
\begin{equation}
    p = x\sqrt{1-\frac{bx^2}{2}}\nonumber
\end{equation}
So, the time period can be obtained from, 
\begin{equation}
    \int^{T/2}_0 d\tau = \int^A_0 \frac{dx}{x\sqrt{1-\frac{bx^2}{2}}}
\end{equation}
Let, $\frac{bx^2}{2} = \cos^2\theta \text{Or, } x = \sqrt{\frac{2}{b}}\cos\theta$. Using this substitution, we have,
\begin{equation}
    \frac{T}{2} = -\int^A_0 \sec\theta d\theta = -\ln|\sec\theta + \tan\theta||_0^A
\end{equation}
Putting, $\theta = \cos^{-1}(\sqrt{\frac{b}{2}}x)$, we obtain,
\begin{equation}
    \frac{T}{2} = -\ln|\frac{2}{b}\frac{1}{x} + \frac{\sqrt{1-x^2}}{x}|_0^A
\end{equation}
As expected, the time period diverges.
\\Now, let $C=E>0$ (the energy of the system). We thus have,
\begin{eqnarray}
    \frac{p^2}{2} - \frac{x^2}{2} + \frac{bx^4}{4} &=& E\nonumber\\
    \text{Or, } \int^{T/2}_0 d\tau &=& \int^A_0 \frac{dx}{x\sqrt{1-\frac{bx^2}{2} + \frac{2E}{x^2}}}
\end{eqnarray}
Making the substitution $x^2 = k$, we change the integral to,
\begin{equation}
    \int^{A^2}_0 \frac{dk}{2\sqrt{k}}\frac{1}{\sqrt{k - \frac{bk^2}{2} + 2E}} = \frac{T}{2}
\end{equation}
This integral, though does not have a closed form, is finite. However, there's a bound to it. Let us check the denominator of the integral a bit more. We have,
\begin{equation}
    \frac{T}{2} = \int^{A^2}_{0^+}\frac{dk}{\sqrt{4k^2 - 2bk^3 + 8Ek}}
\end{equation}
Since this must be a real finite value, the expression within the square root must be greater than zero. We have, thus,
\begin{equation}
    k(4k - 2bk^2 + 8E) > 0
\end{equation}
So there're three zeroes of this cubic expression. One value is of course zero, and from the quadratic expression, we have two roots. These two roots are, viz., $k_1$ and $k_2$ as,
\begin{equation}
    k = \frac{-1 \pm \sqrt{1+4bE}}{b}
\end{equation}
Clearly, $k_1 = \sqrt{\frac{2m\omega}{\hbar}}$ is positive, and $k_2 = \sqrt{i\frac{2}{m\omega\hbar}}$ is negative, since, $\sqrt{1+4bE}>1$. Since, b and E are both positive finite constant quantities, we have these two roots to always exist regardless of these values. Since, k is the square of the displacement variable, it must be positive too. Now, this expression is a globally decreasing function, we'll have the allowed value of the integral for the interval $(0, k_1)$ only, for k. Thus, the integral will only exist for values of $A^2$, for which, 
\begin{equation}
    A^2 < \frac{-1+\sqrt{1+4bE}}{b}
\end{equation}
We get a more explicit form of the bound of $A$ by putting the value of b in CM and QM counterparts, explicitly. 
\section{Conclusion}
We derive the equation of motion of a quantum quartic oscillator in a quasi-classical basis which mimics exactly its classical counterpart, where the coefficients of linear and cubic terms in classical and quantum cases differ significantly. By appropiate scaling of time and rescaling the time period we determine the frequency of the system using Lindstet-Poincare perturbation theory by removing secular terms. The scheme allows us to capture the amplitude dependence of frequency and quantum corrections order by order. An interesting consequence of the analysis was to show that there exist specific values of the nonlinear coefficient for which the frequency becomes independent of amplitude resulting in a condition for isochronicity of the oscillator. We have examined the motion along the separatrix and work out the amplitude dependence and quantum corrections for the bounds for periodicity of the oscillation.
\\Before leaving, we mention that a number of approaches to understand the classical and quantum theory of motion are in order, \cite{article22} \cite{article31}; the present approach based on quasi-classical states of harmonic oscillator specifically suits the purpose since the classical divergence-free perturbation theory is readily applicable without any difficulty. The form of the equation of motion is same in both cases although their contents differ due to its distinct characteristic nature of the coefficients in classical and quantum cases. Although the scheme uses an oscillator with specific nonlinearity, it can be implemented in the case of other non-linear systems as well. The experimental verification and refinement of the bounds and isochronous frequencies found in this article can be done using quantum optical systems.     
\section{Acknowledgements}
The authors report no notable conflict of interest. One of the authors, MB, is also indebted to the lectures he attended delivered by Professor Krishnendu Sengupta in his course curriculum of Advanced Quantum Mechanics.  MB would also acknowledge useful discussions with Shihabul Haque and Nilanjan Sasmal. MB is funded by the Institute fellowship. 
\bibliography{QAO}

%apsrmp4-2.bst 2018-12-27 (MD) hand-edited version of apsrmp4-1.bst
%Control: author (3) reversed first dotless
\begin{thebibliography}{8}%
\makeatletter
\providecommand \@ifxundefined [1]{%
 \@ifx{#1\undefined}
}%
\providecommand \@ifnum [1]{%
 \ifnum #1\expandafter \@firstoftwo
 \else \expandafter \@secondoftwo
 \fi
}%
\providecommand \@ifx [1]{%
 \ifx #1\expandafter \@firstoftwo
 \else \expandafter \@secondoftwo
 \fi
}%
\providecommand \natexlab [1]{#1}%
\providecommand \enquote  [1]{``#1''}%
\providecommand \bibnamefont  [1]{#1}%
\providecommand \bibfnamefont [1]{#1}%
\providecommand \citenamefont [1]{#1}%
\providecommand \href@noop [0]{\@secondoftwo}%
\providecommand \href [0]{\begingroup \@sanitize@url \@href}%
\providecommand \@href[1]{\@@startlink{#1}\@@href}%
\providecommand \@@href[1]{\endgroup#1\@@endlink}%
\providecommand \@sanitize@url [0]{\catcode `\\12\catcode `\$12\catcode `\&12\catcode `\#12\catcode `\^12\catcode `\_12\catcode `\%12\relax}%
\providecommand \@@startlink[1]{}%
\providecommand \@@endlink[0]{}%
\providecommand \url  [0]{\begingroup\@sanitize@url \@url }%
\providecommand \@url [1]{\endgroup\@href {#1}{\urlprefix }}%
\providecommand \urlprefix  [0]{URL }%
\providecommand \Eprint [0]{\href }%
\providecommand \doibase [0]{https://doi.org/}%
\providecommand \selectlanguage [0]{\@gobble}%
\providecommand \bibinfo  [0]{\@secondoftwo}%
\providecommand \bibfield  [0]{\@secondoftwo}%
\providecommand \translation [1]{[#1]}%
\providecommand \BibitemOpen [0]{}%
\providecommand \bibitemStop [0]{}%
\providecommand \bibitemNoStop [0]{.\EOS\space}%
\providecommand \EOS [0]{\spacefactor3000\relax}%
\providecommand \BibitemShut  [1]{\csname bibitem#1\endcsname}%
\let\auto@bib@innerbib\@empty
%</preamble>
\bibitem [{\citenamefont {Akande}\ \emph {et~al.}(2022)\citenamefont {Akande}, \citenamefont {Adjaï}, \citenamefont {Nonti},\ and\ \citenamefont {Monsia}}]{Akande22}%
  \BibitemOpen
  \bibfield  {author} {\bibinfo {author} {\bibnamefont {Akande}, \bibfnamefont {Jean}}, \bibinfo {author} {\bibfnamefont {Kolawolé Kêgnidé~Damien}\ \bibnamefont {Adjaï}}, \bibinfo {author} {\bibfnamefont {Marcellin}\ \bibnamefont {Nonti}}, and\ \bibinfo {author} {\bibfnamefont {Marc~Delphin}\ \bibnamefont {Monsia}}} (\bibinfo {year} {2022}),\ \bibfield  {title} {\enquote {\bibinfo {title} {Isochronous oscillations of nonlinear systems},}\ }in\ \href {https://doi.org/10.5772/intechopen.106354} {\emph {\bibinfo {booktitle} {Nonlinear Systems}}},\ \bibinfo {editor} {edited by\ \bibinfo {editor} {\bibfnamefont {Bo}~\bibnamefont {Yang}}\ and\ \bibinfo {editor} {\bibfnamefont {Dušan}\ \bibnamefont {Stipanović}}},\ Chap.~\bibinfo {chapter} {8}\ (\bibinfo  {publisher} {IntechOpen},\ \bibinfo {address} {Rijeka})\BibitemShut {NoStop}%
\bibitem [{\citenamefont {Eleonskii}\ \emph {et~al.}(1997)\citenamefont {Eleonskii}, \citenamefont {Korolev},\ and\ \citenamefont {Kulagin}}]{Schro}%
  \BibitemOpen
  \bibfield  {author} {\bibinfo {author} {\bibnamefont {Eleonskii}, \bibfnamefont {V}}, \bibinfo {author} {\bibfnamefont {V.}~\bibnamefont {Korolev}}, and\ \bibinfo {author} {\bibfnamefont {N.}~\bibnamefont {Kulagin}}} (\bibinfo {year} {1997}),\ \bibfield  {title} {\enquote {\bibinfo {title} {On a classical analog of the isospectral schrödinger problem},}\ }\href {https://doi.org/10.1134/1.567442} {\bibfield  {journal} {\bibinfo  {journal} {JETP Letters}\ }\textbf {\bibinfo {volume} {65}},\ \bibinfo {pages} {889--893}}\BibitemShut {NoStop}%
\bibitem [{\citenamefont {Filipponi}\ and\ \citenamefont {Cavicchia}(2011)}]{10.1119/1.3579129}%
  \BibitemOpen
  \bibfield  {author} {\bibinfo {author} {\bibnamefont {Filipponi}, \bibfnamefont {A}}, and\ \bibinfo {author} {\bibfnamefont {D.~R.}\ \bibnamefont {Cavicchia}}} (\bibinfo {year} {2011}),\ \bibfield  {title} {\enquote {\bibinfo {title} {{Anharmonic dynamics of a mass O-spring oscillator}},}\ }\href {https://doi.org/10.1119/1.3579129} {\bibfield  {journal} {\bibinfo  {journal} {American Journal of Physics}\ }\textbf {\bibinfo {volume} {79}}~(\bibinfo {number} {7}),\ \bibinfo {pages} {730--735}},\ \Eprint {https://arxiv.org/abs/https://pubs.aip.org/aapt/ajp/article-pdf/79/7/730/13066743/730\_1\_online.pdf} {https://pubs.aip.org/aapt/ajp/article-pdf/79/7/730/13066743/730\_1\_online.pdf} \BibitemShut {NoStop}%
\bibitem [{\citenamefont {Klauder}\ and\ \citenamefont {Skagerstam}(1985)}]{doi:10.1142/0096}%
  \BibitemOpen
  \bibfield  {author} {\bibinfo {author} {\bibnamefont {Klauder}, \bibfnamefont {J}}, and\ \bibinfo {author} {\bibfnamefont {B}~\bibnamefont {Skagerstam}}} (\bibinfo {year} {1985}),\ \href {https://doi.org/10.1142/0096} {\emph {\bibinfo {title} {Coherent States}}}\ (\bibinfo  {publisher} {WORLD SCIENTIFIC})\ \Eprint {https://arxiv.org/abs/https://www.worldscientific.com/doi/pdf/10.1142/0096} {https://www.worldscientific.com/doi/pdf/10.1142/0096} \BibitemShut {NoStop}%
\bibitem [{\citenamefont {Rand}(2012)}]{article}%
  \BibitemOpen
  \bibfield  {author} {\bibinfo {author} {\bibnamefont {Rand}, \bibfnamefont {Richard}}} (\bibinfo {year} {2012}),\ \bibfield  {title} {\enquote {\bibinfo {title} {Lecture notes on nonlinear vibrations},}\ }\href {https://audiophile.tam.cornell.edu/randpdf/nlvibe52.pdf} {\ }\BibitemShut {NoStop}%
\bibitem [{\citenamefont {Saha}\ and\ \citenamefont {Gangopadhyay}(2017)}]{article22}%
  \BibitemOpen
  \bibfield  {author} {\bibinfo {author} {\bibnamefont {Saha}, \bibfnamefont {Sandip}}, and\ \bibinfo {author} {\bibfnamefont {Gautam}\ \bibnamefont {Gangopadhyay}}} (\bibinfo {year} {2017}),\ \bibfield  {title} {\enquote {\bibinfo {title} {Isochronicity and limit cycle oscillation in chemical systems},}\ }\href {https://doi.org/10.1007/s10910-016-0729-1} {\bibfield  {journal} {\bibinfo  {journal} {Journal of Mathematical Chemistry}\ }\textbf {\bibinfo {volume} {55}},\ \bibinfo {pages} {887--910}}\BibitemShut {NoStop}%
\bibitem [{\citenamefont {Sarkar}\ and\ \citenamefont {Bhattacharjee}(2012)}]{article31}%
  \BibitemOpen
  \bibfield  {author} {\bibinfo {author} {\bibnamefont {Sarkar}, \bibfnamefont {A}}, and\ \bibinfo {author} {\bibfnamefont {J.~K.}\ \bibnamefont {Bhattacharjee}}} (\bibinfo {year} {2012}),\ \bibfield  {title} {\enquote {\bibinfo {title} {Renormalisation group and isochronous oscillations},}\ }\href {https://doi.org/https://doi.org/10.1140/epjd/e2012-20427-8} {\bibfield  {journal} {\bibinfo  {journal} {The European Physical Journal D}\ }\textbf {\bibinfo {volume} {66}},\ https://doi.org/10.1140/epjd/e2012-20427-8}\BibitemShut {NoStop}%
\bibitem [{\citenamefont {Sharatchandra}(1997)}]{sharatchandra1997coherentstatesanharmonicoscillator}%
  \BibitemOpen
  \bibfield  {author} {\bibinfo {author} {\bibnamefont {Sharatchandra}, \bibfnamefont {H~S}}} (\bibinfo {year} {1997}),\ \href {https://arxiv.org/abs/quant-ph/9707032} {\enquote {\bibinfo {title} {Coherent states for the anharmonic oscillator and classical phase space trajectories},}\ }\Eprint {https://arxiv.org/abs/quant-ph/9707032} {arXiv:quant-ph/9707032 [quant-ph]} \BibitemShut {NoStop}%
\end{thebibliography}%

\end{document}